\documentclass{webofc}

\usepackage[varg]{txfonts}   
\usepackage{hyperref}
\usepackage{url}
\usepackage{color}

\hypersetup{colorlinks=true,citecolor=blue,urlcolor=blue,linkcolor=blue}
\begin{document}
\title{A Deep Learning Based Estimator for Light Flavour Elliptic Flow in Heavy Ion Collisions at  LHC Energies}
%
%

\author{\firstname{Gergely G\'abor} \lastname{Barnaf\"oldi}\inst{1}\fnsep\thanks{\email{Barnafoldi.Gergely@wigner.hun-ren.hu}} \and
        \firstname{Neelkamal} \lastname{Mallick}\inst{1,2,3}
        \and
        \firstname{Suraj} \lastname{Prasad}\inst{1,2}
        \and
        \firstname{Raghunath} \lastname{Sahoo}\inst{2}
        \and
        \firstname{Aditya} \lastname{Nath Mishra}\inst{4}
}
\institute{HUN-REN Wigner Research Center for Physics, P.O. Box 49, 1125 Budapest, Hungary
\and     
Department of Physics, Indian Institute of Technology Indore, Simrol, Indore 453552, India
\and 
University of Jyv\"askyl\"a, Department of Physics, P.O. Box 35, FI-40014, Jyv\"askyl\"a, Finland
\and
University Centre of Research and Development Department, Chandigarh University, Gharuan, Mohali-140413, Punjab, India
}

\abstract{We developed a deep learning feed-forward network for estimating elliptic flow ($v_2$) coefficients in heavy-ion collisions from RHIC to LHC energies. The success of our model is mainly the estimation of $v_2$ from final state particle kinematic information and learning the centrality and the transverse momentum ($p_{\rm T}$) dependence of $v_2$ in wide $p_{\rm T}$ regime. The deep learning model is trained with AMPT-generated Pb-Pb collisions at $\sqrt{s_{\rm NN}} = 5.02$ TeV minimum bias events. We present $v_2$ estimates for $\pi^{\pm}$, $\rm K^{\pm}$, and $\rm p+\bar{p}$ in heavy-ion collisions at various LHC energies. These results are compared with the available experimental data wherever possible.
}
\maketitle
\section{Introduction}
\label{sec:intro}

Ultrarelativistic heavy-ion collisions have been studied extensively for decades in experiments at the Relativistic Heavy Ion Collider (RHIC BNL) and at the Large Hadron Collider (LHC CERN). In these collisions, a deconfined thermalized medium of quarks and gluons can be formed~\cite{Bass:1998vz}. This medium of hot and dense state of the strongly interacting matter is called the quark-gluon plasma (QGP). Only signatures of the formation of QGP could be studied using various indirect effects such as jet quenching, strangeness enhancement, and quarkonia suppression since no direct observation is possible due to the short lifetime of the strongly interacting matter.

Transverse collective flow is another key observable, which is widely studied to investigate the properties of QGP in heavy-ion collisions~\cite{Heinz:2013th}. This is anisotropic and depends on the equation of state and transport coefficients of the system. Anisotropic flow signifies the formation of QGP medium in noncentral relativistic heavy-ion collisions. The pressure gradient formed in the hot and dense medium due to the initial spatial anisotropy can transform into final state momentum space azimuthal anisotropy. This momentum anisotropy could be expressed as the coefficients of the Fourier expansion of the azimuthal momentum distribution of the produced particles. The second-order flow coefficient is the so-called elliptic flow ($v_2$). Finite azimuthal anisotropy has been well observed in heavy-ion collision experiments so far at RHIC and LHC energies up to higher-order cumulants with various analysis methods~\cite{STAR:2003wqp,ALICE:2010suc,ALICE:2011ab,ALICE:2014dwt}. Here, we present our deep learning feed-forward network for estimating elliptic flow ($v_2$) coefficients, which we compare to heavy-ion collision data from RHIC to LHC energies.

\section{The Model and the DNN architecture}

Anisotropic flow can be measured and quantified by the coefficients of Fourier expansion of the azimuthal momentum distribution, given by~\cite{Voloshin:1994mz}: 
\begin{equation}
        \frac{\mathrm{d}N}{\mathrm{d}\phi}=\frac{1}{2\pi}\left(1+2\sum^{\infty}_{n=1}v_{n}\cos \left[n(\phi-\psi_{n})\right]\right) \ \ \mathrm{with} \ \ v_{\rm n}=\langle\cos[n(\phi-\psi_{\rm n})]\rangle \ ,
    \label{eqn-invfourierexp}
\end{equation}
where, $v_{\rm n}$ denotes $n ^{\rm th}$ order anisotropic flow coefficient, $\phi$ is the azimuthal angle, and $\psi_{\rm n}$ is the corresponding harmonic symmetry plane angle. 
In order to calculate the elliptic flow event-by-event, we have used the event plane method~\cite{Masera:2009zz}, and for simplicity, we have fixed the reaction plane angle, $\psi_{\rm R} = 0$, which results in $v_{\rm 2} = \langle\cos(2\phi)\rangle$.

A deep learning-based machine learning algorithm was developed to estimate the elliptic flow event-by-event. For training the deep neural network (DNN), we have used a multiphase transport (AMPT) model to simulate the dataset. AMPT is a Monte Carlo-based event simulator that is used to generate ultrarelativistic nucleus-nucleus collisions at RHIC and LHC energies~\cite{Lin:2004en}. AMPT has four components, namely, initialization of collisions by HIJING~\cite{ampthijing}, parton transport by Zhang's Parton Cascade model~\cite{amptzpc}, hadronization of the partons performed by spatial coalescence mechanism in string melting mode and Lund string fragmentation model in the default version of AMPT~\cite{Lin:2001zk,He:2017tla}, and finally, the hadron transport using a relativistic transport model~\cite{amptart1,amptart2}. The DNN was trained with Pb-Pb collisions at $\sqrt{s_{\rm NN}} = 5.02$ TeV minimum bias events with all charged particles having $p_{\rm T}> 0.15$ GeV/c in pseudorapidity, $|\eta|<0.8$.
%
\begin{figure}[h]
\centering
\includegraphics[width=12cm,clip]{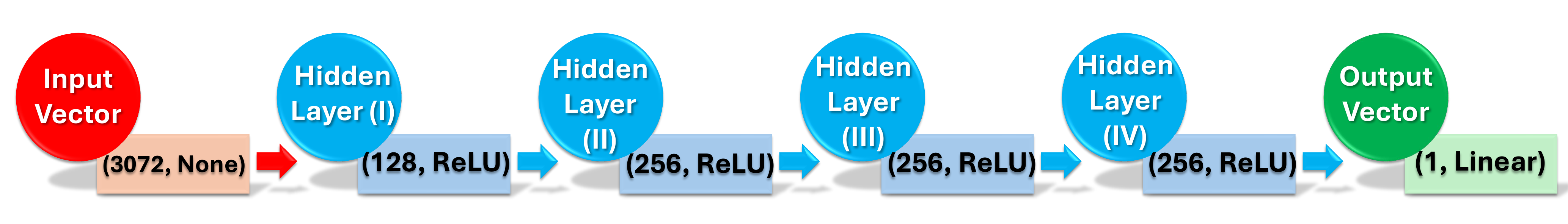} 
\caption{The structure of the DNN architecture used for the $v_2$ estimation. The number of
nodes and the type of activation function used in each layer are denoted.}
\label{fig:dnnstruc}    
\end{figure}

For this regression problem, the DNN consists of one input layer, four hidden layers, and one output layer, as depicted in Fig.~\ref{fig:dnnstruc}. The inputs are given as "pictures" with pixels of normalized transverse momenta, mass and energy values on the pseudorapidity-azimuthal plane. This input with 3072 features are mapped to the first dense layer with 128 nodes, which is connected to the output layer via three hidden layers in succession, each having 256 nodes. The dense layers use the rectified linear unit as the activation function, and the output layer has a single node with a linear activation function. The DNN model uses the \textit{adam} optimizer with \textit{mean squared error} loss function. Details can be found in Refs.~\cite{Mallick:2022alr,Mallick:2023vgi}.

\section{Comparing DNN predictions to AMPT and to experimental data}
\label{sec:results}
\begin{figure*}
\centering
\vspace*{1cm}       
\includegraphics[width=10cm,clip]{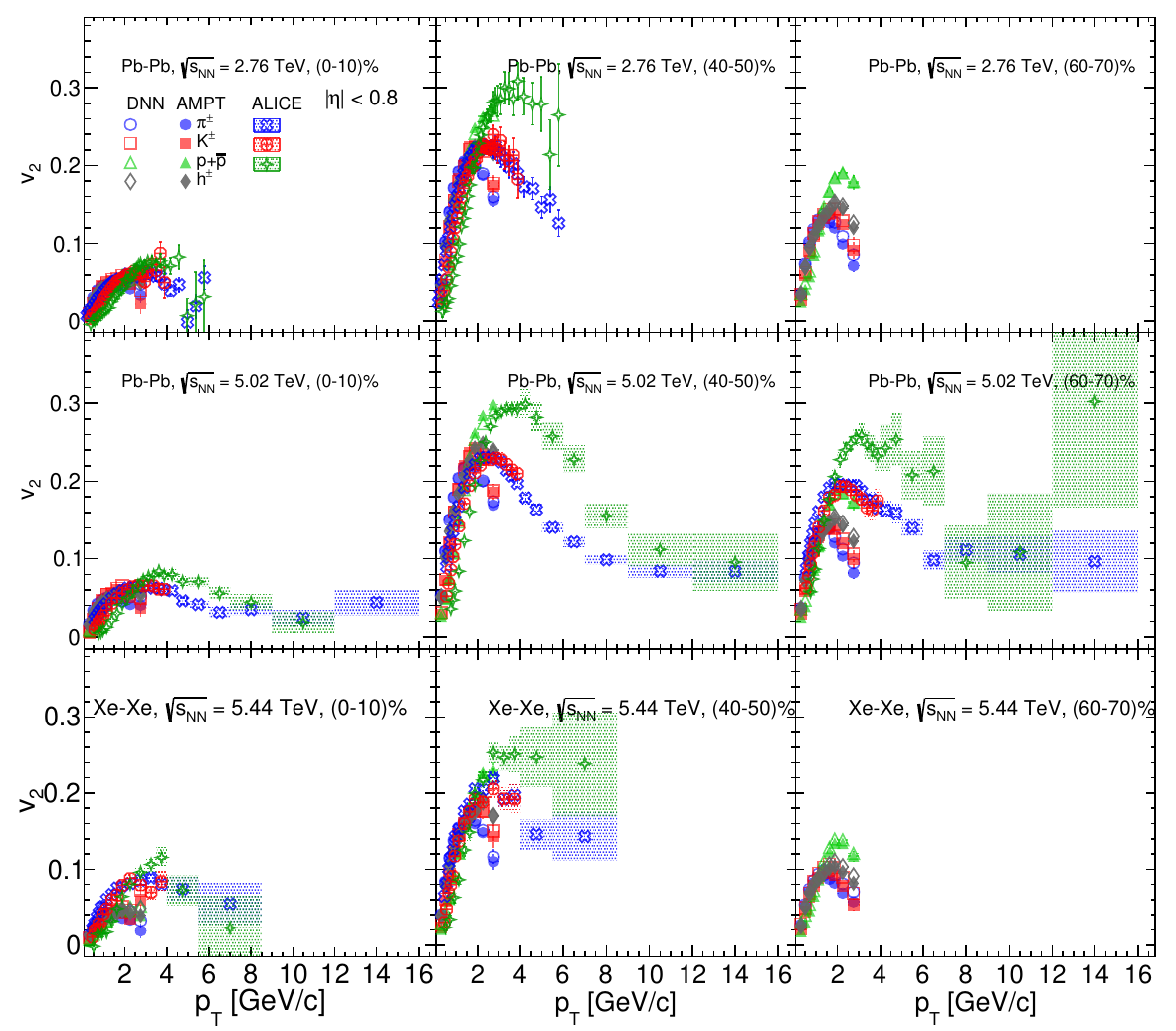}
\caption{Centrality dependence of $v_2({p_{\rm T})}$ for $\pi^{\pm}$, $\rm K^{\pm}$, and $\rm p+\bar{p}$, and all charged hadrons ($h^{\pm}$). The values from AMPT and the predictions from DNN are shown in comparison with ALICE data~\cite{ALICE:2018yph}.}
\label{fig:2}       
\end{figure*}
The elliptic flow, $v_2(p_{\rm T})$ for identified hadrons: $\pi^{\pm}$, $\rm K^{\pm}$, and $\rm p+\bar{p}$ in nucleus-nucleus collisions at $\sqrt{s_{\rm NN}} = 2.76$~TeV (Pb-Pb), 5.02~TeV (Pb-Pb), and 5.44~TeV (Xe-Xe) is plotted in Fig.~\ref{fig:2} from top to bottom, respectively. To be comparable with ALICE data~\cite{ALICE:2018yph}, the AMPT and DNN were simulated with tracks with $p_{\rm T}> 0.5$ GeV/c and in midrapidity, $|y|<0.5$. Three selected collision centrality ranges were used for the plots: 0-10\%, 40-50\% and 60-70\%. 

One can see from the results, that all particle species behave similarly: the magnitude of $v_2(p_{\rm T})$ increases with increasing $p_{\rm T}$ until it reaches a maximum value around $p_{\rm T} \approx 2.0$ GeV/c, and then it starts to decrease beyond this point. The values of $v_2(p_{\rm T})$ from AMPT obtained in this region (\textit{i.e.} $p_{\rm T} \lesssim 2.0\rm{-}3.0$ GeV/c) is comparable in magnitude with ALICE results for the individual particle cases. However, beyond this $p_{\rm T}$ value, AMPT fails to describe the data as $v_2(p_{\rm T})$ falls faster with increasing $p_{\rm T}$ since fragmentation takes over from coalescence at high~$p_{\rm T}$.

DNN predictions agree with AMPT values quite nicely up to $p_{\rm T}\lesssim 4.0\rm{-}6.0$ GeV/c. Beyond this $p_{\rm T}$, the values from DNN start to differ from the AMPT calculated data points. The reason behind that is the statistics, which decrease for the higher $p_{\rm T} \gtrsim 6.0$ GeV/c regions. The limited number of event provided less instances to the DNN model during the training process. For this reason, the mismatch between DNN and AMPT comes into the picture beyond $p_{\rm T} \gtrsim 6.0$ GeV/c.

\section{Discussion and Summary}

In Refs.~\cite{Mallick:2022alr,Mallick:2023vgi}, we demonstrated the applicability of a DNN-based machine learning model to evaluate the second-order anisotropic flow coefficient ($v_2$) event-by-event for identified hadrons from final state particle kinematic information in heavy-ion collisions. 
The developed DNN model can well estimate $v_2$ for light-flavor identified particles such as $\pi^{\pm}$, $\rm K^{\pm}$, and $\rm p+\bar{p}$ in heavy-ion collisions at RHIC and LHC energies. Here, we compared the results to the data. The DNN was trained with AMPT data of minimum bias Pb-Pb collisions at $\sqrt{s_{\rm NN}}$ = 5.02 TeV and was able to learn and predict the centrality, hadron flavor, energy and transverse momentum dependence of elliptic flow for other collision systems at various energies. 
Results were presented for Pb-Pb collisions at $\sqrt{s_{\rm NN}}$ = 2.76 TeV, Xe-Xe collisions at $\sqrt{s_{\rm NN}}$ = 5.44 TeV in three centrality bins. We have seen DNN estimator and the AMPT data correlate well up to $p_{\rm T} \lesssim 3$~GeV/$c$, where the training statistics issue vanishes. Comparison with available ALICE dataset~\cite{ALICE:2018yph} at LHC energies follows the trends of the AMPT with a high accuracy, which latter underestimates the data above $p_{\rm T} \approx 2$~GeV/$c$, similarly as the DNN model. The obtained results suggest, that original data or Monte Carlo simulations with better agreement with data at high $p_{\rm T}$, can train the DNN-estimator for more accurate predictions.

\section*{Acknowledgements}
SP acknowledges the doctoral fellowship from UGC, Govt. of India. NM, SP and RS acknowledge the DAE-DST, Govt. of India funding under the mega-science project – “Indian participation in the ALICE experiment at CERN” bearing Project No. SR/MF/PS-02/2021-IITI (E-37123). GGB acknowledges the Hungarian National Research, Development and Innovation Office (NKFIH) under Contract No. OTKA K135515, 2021-4.1.2-NEMZ\_KI-2024-00031, 2024-1.2.5-TÉT-2024-00022; Wigner Scientific Computing Laboratory (WSCLAB). The MoU between IIT Indore and HUN-REN Wigner RCP, Hungary, for the techno-scientific cooperation is highly appreciated. ANM would like to thank UCRD, Chandigarh University for their research facilities.

%

\begin{thebibliography}{100}

\bibitem{Bass:1998vz}
S.~A.~Bass, M.~Gyulassy, H.~Stoecker and W.~Greiner,
J. Phys. G \textbf{25}, R1-R57 (1999). 

\bibitem{Heinz:2013th}
U.~Heinz and R.~Snellings,
Ann. Rev. Nucl. Part. Sci. \textbf{63}, 123 (2013).

\bibitem{STAR:2003wqp}
J.~Adams \textit{et al.} [STAR Collaboration],
Phys. Rev. Lett. \textbf{92}, 052302 (2004).

\bibitem{ALICE:2010suc}
K.~Aamodt \textit{et al.} [ALICE Collaboration],
Phys. Rev. Lett. \textbf{105}, 252302 (2010).

\bibitem{ALICE:2011ab}
K.~Aamodt \textit{et al.} [ALICE Collaboration],
Phys. Rev. Lett. \textbf{107}, 032301 (2011).

\bibitem{ALICE:2014dwt}
B.~B.~Abelev \textit{et al.} [ALICE Collaboration],
Phys. Rev. C \textbf{90}, 054901 (2014).

\bibitem{Voloshin:1994mz}
S.~Voloshin and Y.~Zhang,
Z. Phys. C \textbf{70}, 665 (1996).

\bibitem{Masera:2009zz}
M.~Masera, G.~Ortona, M.~G.~Poghosyan and F.~Prino,
Phys. Rev. C \textbf{79}, 064909 (2009).

\bibitem{Lin:2004en}
Z.~W.~Lin, C.~M.~Ko, B.~A.~Li, B.~Zhang and S.~Pal,
Phys. Rev. C \textbf{72}, 064901 (2005).

\bibitem{ampthijing}
X.~N.~Wang and M.~Gyulassy,
Phys.\ Rev.\ D {\bf 44}, 3501 (1991).

\bibitem{amptzpc}
B.~Zhang, \ Comput. \ Phys. \ Commun. {\bf 109}, 193 (1998).


\bibitem{Lin:2001zk}
Z.~w.~Lin and C.~M.~Ko,
Phys. Rev. C \textbf{65}, 034904 (2002).

\bibitem{He:2017tla}
Y.~He and Z.~W.~Lin,
Phys. Rev. C \textbf{96}, 014910 (2017).

\bibitem{amptart1}
B.~Li, A.~T.~Sustich, B.~Zhang and C.~M.~Ko,
Int.\ J.\ Mod.\ Phys.\ E {\bf 10}, 267 (2001).

\bibitem{amptart2}
B.~A.~Li and C.~M.~Ko,
Phys.\ Rev.\ C {\bf 52}, 2037 (1995).


\bibitem{Mallick:2022alr}
N.~Mallick  \textit{et al.}, 
Phys. Rev. D \textbf{105}, 114022 (2022).


\bibitem{Mallick:2023vgi}
N.~Mallick \textit{et al.}, 
Phys. Rev. D \textbf{107}, 094001 (2023).



\bibitem{ALICE:2018yph}
S.~Acharya \textit{et al.} [ALICE Collaboration],
JHEP \textbf{09}, 006 (2018).




























































\end{thebibliography}
%
%


\end{document}